\documentclass[11pt,twoside]{article}

\usepackage{asp2006}
\usepackage{epsf}
\usepackage{psfig}
\usepackage{lscape}

\begin{document}
\title{The Structure and Content of Galaxy Outskirts}   
\author{Annette M. N. Ferguson}   
\affil{Institute for Astronomy, University of Edinburgh, Blackford Hill, Edinburgh UK EH9 3HJ}    
\begin{abstract} 
  The outer regions of galaxies are expected to contain important
  clues about the way in which galaxies are assembled.  Although
  quantitative study of these parts has been severely limited in the
  past, breakthroughs are now being made thanks to the combination of
  wide-area star counts, deep HST imagery and 8-m class spectroscopy.
  I review here several recent results concerning substructure, star
  clusters and stellar halos in the outer regions of our nearest large
  neighbours, M31 and M33.
\end{abstract}

\section{The Outskirts of  Galaxies: Motivation and History}

The study of the faint outskirts of galaxies has become increasingly
important in recent years. From a theoretical perspective, it has been
realised that many important clues about the galaxy assembly process
should lie buried in these parts. Cosmological simulations of disk
galaxy formation incorporating baryons now yield predictions for the
large-scale structure and stellar content at large radii -- for
example, the abundance and nature of stellar substructure and the
ubiquity, structure and content of stellar halos and thick
disks. These models generally predict a wealth of (sub)structure at
levels of $\mu_V\sim 30$ mag/$\sq''$ and fainter (e.g. Bullock \&
Johnson 2005); their verification thus requires imagery and
spectroscopy of galaxies to ultra-faint surface brightness levels.

Since Malin first applied his photographic stacking and amplification
technique (e.g. Malin et al. 1983), it has been known that some
galaxies possess unusual low surface brightness (LSB) structures --
shells, loops, asymmetric envelopes -- in their outer regions.
Although limited to $\mu_B\la 28$ mag/$\sq''$, these images were
sufficient to demonstrate that even the most ``normal'' nearby
galaxies could become very abnormal when viewed at faint light levels
(e.g. Weil et al 1997). Follow-up study of Malin's LSB features, and
several more recently-discovered examples (e.g. Shang et al. 1999),
has been severely limited due to the technical difficulties associated
with detecting and quantifying diffuse light at surface brightnesses
$\sim10$ magnitudes below sky.  The currently most viable technique to
probe the very low surface brightness regions of galaxies is that of
wide-area resolved star counts.  I review here recent results from
studies of the resolved populations in the outer regions of M31 and
M33, focusing on substructure, star clusters and stellar halos.

\section{Stellar Substructure in the Outskirts of  Galaxies}

Figure 1 (left panel) shows a map of the red giant branch (RGB)
population (age $\ga 1-2$~Gyr) in M31 made with the INT Wide-Field
Camera (see Ferguson et al. 2002 and Irwin et al. 2005 for details);
the right panel indicates regions of prominent stellar substructure in
the inner halo.  The faintest features visible by eye in the M31 map
have effective V-band surface brightnesses of $\mu_V\sim28-31$
mag/$\sq''$.  An ongoing survey with Megacam on CFHT extends the INT
coverage in the south-east quadrant of M31 to a projected radius of
$\sim 150$~kpc and has already led to the discovery of three new dwarf
spheroidal galaxies and the most remote currently known M31 globular
cluster (Martin et al. 2006).  A map of the RGB population around the
low mass spiral M33 has also been made.  Despite reaching the same
limiting depth ($\approx 3$ magnitudes below the tip of the RGB) as
the M31 map, no visible substructure can be discerned around M33
(Ferguson et al. 2007, in prep) ; if substructure is present in this
system, it must be of significantly lower surface brightness.

\begin{figure}
\centerline{\psfig{file=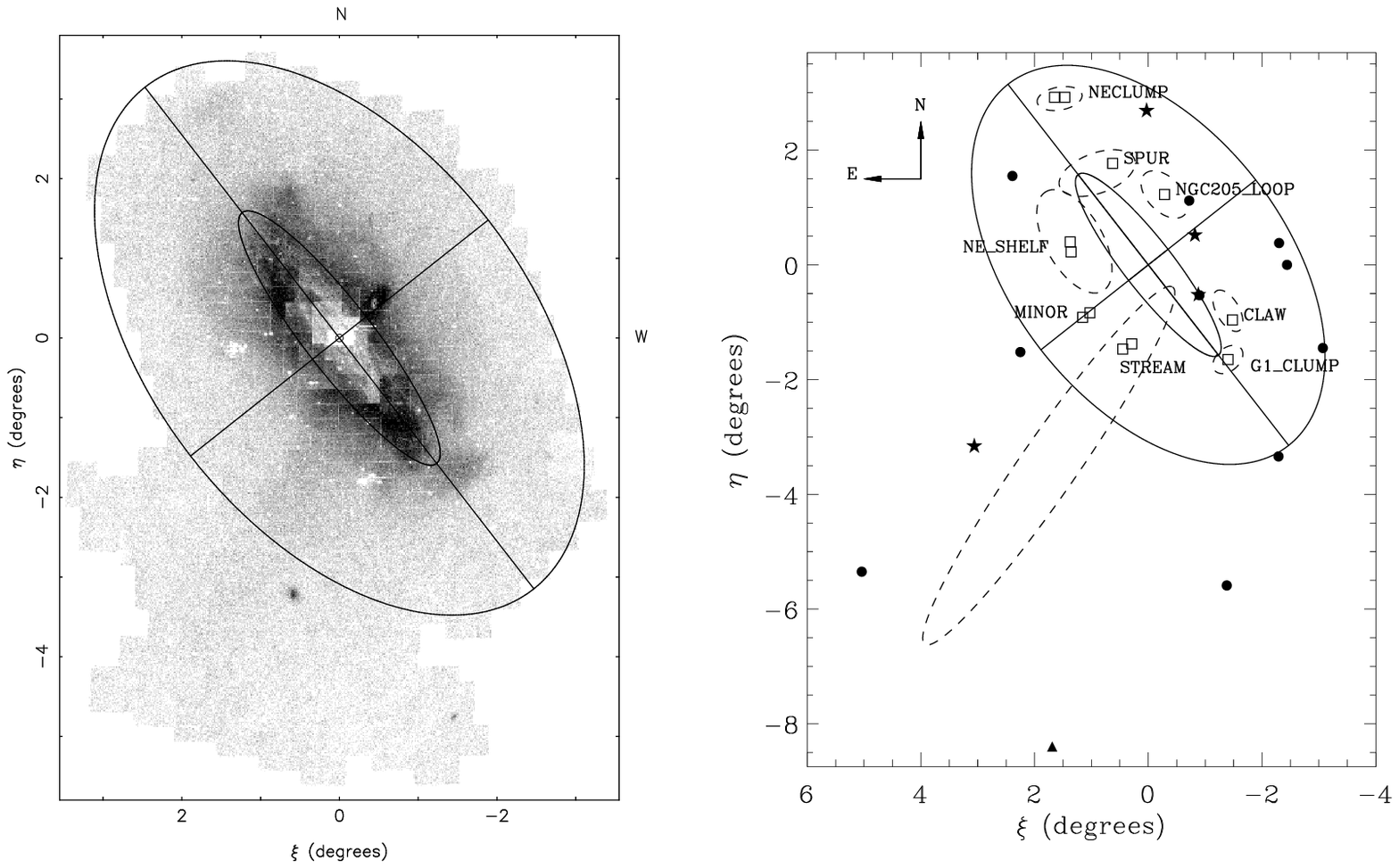,height=9cm,clip=}}
\caption{The INT/WFC RGB star count map of M31 (left) with regions of
prominent stellar substructure identified (right). The star count map
spans $125\times95$~kpc$^2$, corresponding to $\sim 40$ square degrees
at the distance of M31.  Small squares in the right-hand panel
indicate the locations of our deep ACS substructure pointings.
Globular clusters studied by Mackey et al. (2006, 2007) are indicated
by stars (extended clusters) and filled circles (classical
clusters). The newly-discovered globular cluster of Martin et
al. (2006) is indicated with a filled triangle.}
\end{figure}

Various distinct features can be seen around M31, including a giant
stream in the south-east, stellar overdensities at large radii along
the major axis, a diffuse structure in the north-east and a loop of
stars projected near NGC~205.  It is of obvious importance to
establish the origin of this substructure.  Bullock \& Johnston (2005)
predict substructure in the outer regions of galaxies due to tidal
debris from the accretion and disruption of an expected population of
$\sim100-200$ luminous satellites.  It has long been recognized
that satellite accretion will also heat and restructure the thin disk,
generating additional ``low latitude'' debris (e.g. Quinn et al. 1993,
Gauthier et al.  (2006).  I summarize here our current understanding
of the substructure seen around M31:

\begin{itemize}

\item{Deep HST/ACS CMDs reaching well below the horizontal branch have
    been obtained for 8 regions in the far outskirts of M31, including
    seven regions of visible substructure (see Figure 1). Many of the
    CMDs exhibit different morphologies, suggesting variations in the
    mean age and metallicity of the constituent stars (Ferguson et
    al. 2005, Richardson et al., these proceedings).  In all cases,
    the metallicity inferred from the RGB colour is significantly
    higher ([Fe/H]$\ga-0.7$) than that of typical present-day low mass
    Local Group dwarf satellites ([Fe/H]$\la-1.5$), implying that such
    objects are unlikely to be the progenitors of the observed
    substructure.  }
\item{All substructure fields lying near the major axis contain stars
    of age $\la 1.5$~Gyr (Faria et al 2007; Richardson et al.,these
    proceedings); this includes the NE clump field, which lies at a
    projected radius of 40~kpc. Additionally, the major axis
    substructure fields are dominated by a strong rotational
    signature, similar to the HI disk, with a modest velocity
    dispersion (Ibata et al. 2005). Taken together, these results
    suggest that the ``low-latitude'' substructure in M31 originates
    primarily from perturbed thin disk and not from an accreted
    satellite (e.g. Pe{\~n}arrubia et al. 2006).}
  \item{The combination of line-of-sight distances and radial
      velocities for stars at various locations along the giant stream
      constrains the progenitor orbit (e.g.  Ibata et al. 2004; Fardal
      et al. 2006).  Currently-favoured orbits do not easily connect
      the more luminous inner satellites (e.g. M32, NGC~205) to the
      stream however this finding leaves some remarkable coincidences
      (e.g. the projected alignment on the sky, similar metallicities)
      yet unexplained.  The giant stream is linked to another
      overdensity, the diffuse feature lying north-east of M31's
      centre, on the basis of nearly identical CMD morphologies and
      RGB luminosity functions (Ferguson et al. 2005); orbit
      calculations suggest this connection is indeed likely. }
\end{itemize}  

\section{Star Clusters}

Our INT and Megacam surveys of the outskirts of M31 and M33 have
facilitated a search for new globular clusters in these regions (Huxor
et al. 2005, Martin et al. 2006, Huxor 2007, Huxor et al. 2007, in
prep).  To date, almost 40 new clusters have been identified in M31
and 5 in M33. Of particular interest has been the discovery of a new
class of extended, luminous globular clusters in M31 (Huxor et
al. 2005). These objects have large half-light radii ($\approx 30$~pc)
yet luminosities (M$_V \approx -7$) which place them near the peak of
the globular cluster luminosity function.  Such a combination of
luminosity and size has rarely been observed within the star cluster
population.  When placed on the M$_V$--R$_h$ plane, these extended
clusters encroach on the gap in parameter space between classical
globular clusters and dwarf spheroidal galaxies (e.g. Belokurov et
al. (2007)).

\begin{figure}
\plottwo{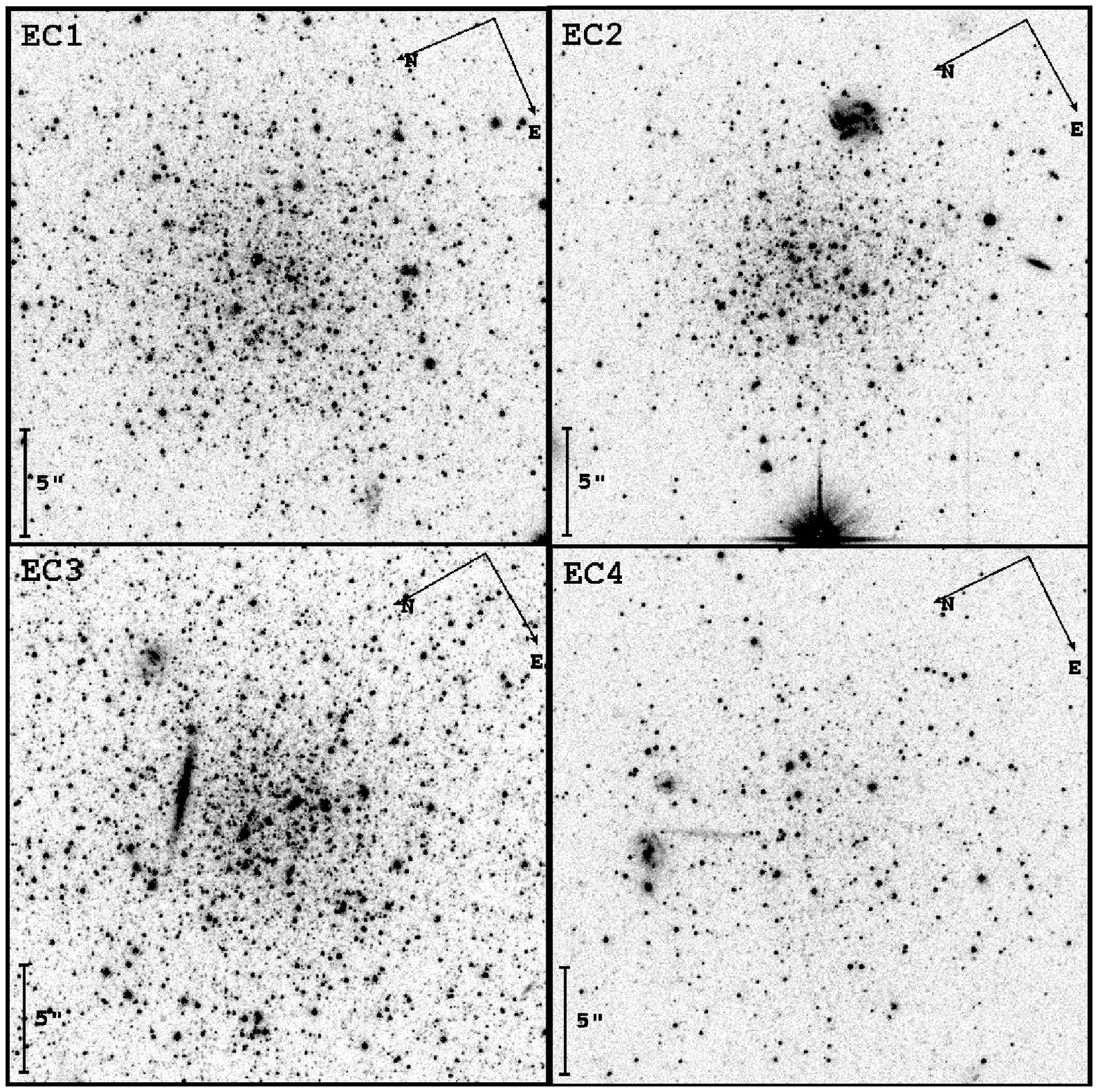}{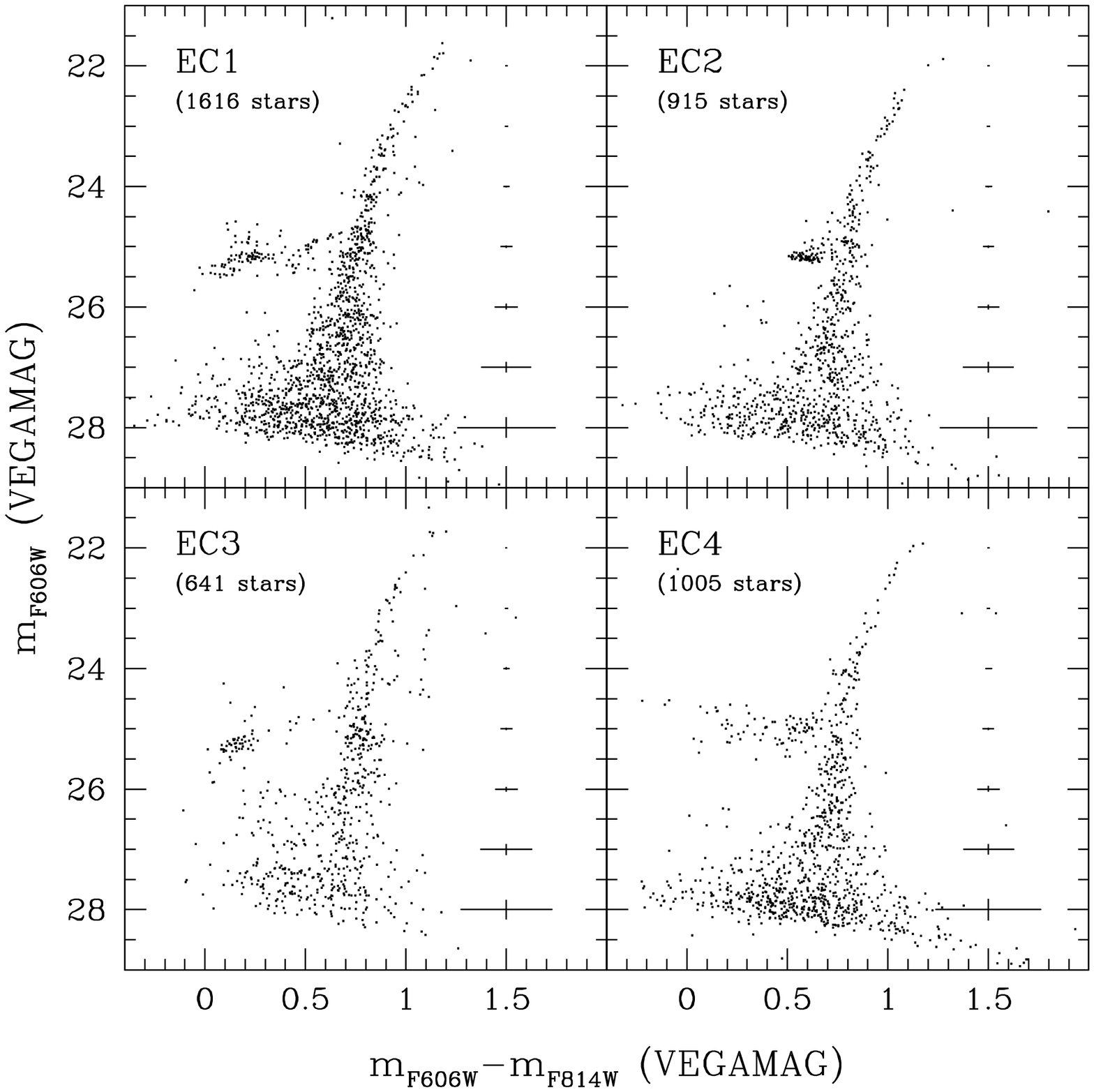}
\caption{Drizzled ACS/WFC F606W images of four extended
  clusters in the outskirts of M31 (left panel).  Each 
image spans 25\arcsec on a side.  The right-hand panel
shows ACS CMDs for these clusters.}
\end{figure}

In order to gain further insight into these objects,
deep HST/ACS observations were obtained for a small sample of extended
clusters lying within 60~kpc of M31 (see Figure 1 and Mackey et
al. 2006).  Figure 2 shows images of these objects and their
associated CMDs.  In all cases, the CMD morphology is described by a
narrow steep RGB and a clear horizontal branch.  Three of the clusters
possess horizontal branches extending to the blue and display
broadened regions at intermediate colour, suggestive of the presence
of RR Lyrae stars.  A second parameter effect can also be discerned
within the sample.

The extended clusters thus appear to be composed of old ($\ge 10$ Gyr)
single stellar populations.  Metallicities, estimated from fitting
Galactic globular cluster fiducials, indicate these objects are also
metal-poor with $-2.2\le$[Fe/H]$\le-1.8$.  Despite their unusual
structures and special location in the M$_V$--R$_h$
plane, these systems appear to be genuine globular clusters as opposed
to the stripped cores of dwarf galaxies.  Studies of their internal
kinematics will be required to test for the presence of dark matter
and thus confirm this interpretation.

We have also obtained deep HST/ACS observations of a sample of
newly-discovered classical globular clusters in the far outskirts of
M31 ($15\le \rm{R_{proj}} \le 100$~kpc, see Figure 1). Figure 3 shows
their CMDs with Galactic globular fiducials overlaid (see also Mackey
et al. 2007).  The bulk of the sample appears old and metal-poor
($-2.2\le$[Fe/H]$\le-1.8$).  The very outermost clusters (R$_{proj}\ge
40$~kpc) are particularly compact (R$_h\approx 5$pc) and
luminous (M$_V\approx-8.5$) when compared to their counterparts in the
outermost halo of the Milky Way.  They are also considerably more
metal-poor than the kinematically-selected field halo population in
M31 (see below), again in contrast to the Milky Way where the field
halo and globular clusters both peak at [Fe/H]$\sim -1.5$.  These
disparities are intriguing and must reflect differences in
the early formation and evolution of the two galaxies or in their
subsequent accretion histories.

\begin{figure}
\centerline{\psfig{file=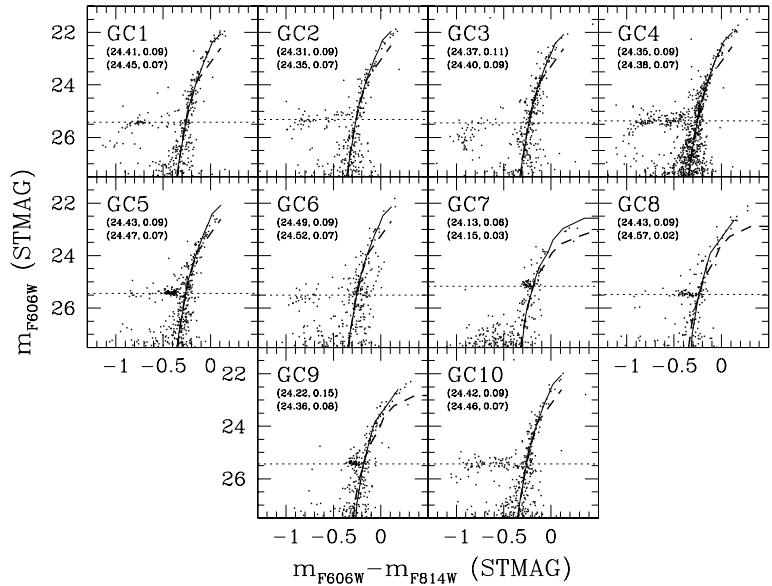,height=8cm}}
\caption{CMDs for 10 newly-discovered classical globular clusters in
the outskirts of M31.  The two best fitting Galactic globular cluster
fiducials are indicated in each case. Note that the CMD of GC7
suggests it is an intermediate-age object. }
\end{figure}

\section{Diffuse Stellar Halos}

Tidal debris from accretion events will disperse over time and streams
from ancient events will have long merged to produce a diffuse stellar
halo (e.g. Abadi et al. 2005; Bullock \& Johnston et al 2005).  If
hierarchical growth is the main mode of mass assembly then such
halos should be a generic feature of galaxies. Unfortunately,
few observational constraints exist on the nature and
ubiquity of stellar halos around galaxies (see however Zibetti et
al. 2004 and Zibetti \& Ferguson 2004).  Even our understanding of the
M31 and M33 stellar halos has been extremely poor until recently. I
summarize some key results here:

\subsection{M31}
Using data from the INT/WFC survey, Irwin et al. (2005) combined
diffuse light surface photometry with resolved star counts to probe
the minor axis profile of M31 to a radius of $\sim55$~kpc (effective
$\mu_V\sim32$~mag/$\sq"$).  The profile shows an unexpected flattening
(relative to the inner R$^{1/4}$ decline) at a radius of $\sim30$~kpc,
beyond which it can be described by shallow power-law (index
$\approx-2.3$), possibly extending out to 150~kpc (Kalirai et
al. 2006).  This compares favourably with the Milky Way halo, which
exhibits a power-law index of $-3.1$ in volume density (Vivas et
al. 2006).  The discovery of a power-law component which dominates the
light at large radii in M31 has profound implications for the
interpretation of all prior studies of the M31 ``stellar halo''. Since
these studies generally targetted regions lying within 30~kpc along
the minor axis, they most likely probed the extended disk/bulge region
of the galaxy and not the true stellar halo.

Keck/DEIMOS spectroscopy has been used to study the kinematics and
metallicities of stars in the far outer regions of M31. By windowing
out the stars which corotate with the HI disk, Chapman et al. (2006)
have detected an underlying metal-poor ([Fe/H]$\sim-1.4$), high
velocity dispersion ($\sigma\sim100$~km/s) component (see also Kalirai
et al. 2006).  Although it has yet to be proven that this component is
the same one that dominates the power-law profile at very large
radius, the evidence is highly suggestive.  Despite previous views to
the contrary, it thus appears that M31 does indeed have a stellar halo
which resembles that of the Milky Way in terms of structure,
metallicity and kinematics.

\subsection{M33}
Studies of the M33 stellar halo have also had a checkered history.
Mould \& Kristian (1986) measured [M/H]$\sim-2.2$ in a field located
at 7~kpc along the minor axis of M33 and, for the better part of two
decades, this metallicity was generally assumed to reflect that of the
M33 halo.  Recent work has found a significantly higher metallicity
for stars in this same field and, at the same time, suggested that the
field is actually dominated by the outer disk and not the stellar halo
(e.g. Tiede et al. 2004, Ferguson et al. 2007 in prep).  The detection
of a power-law structural component in the outskirts of M33 has so far
proved elusive, although the RGB clearly becomes narrower and more
metal-poor in these parts (Ferguson et al. 2007 in prep).  A recent
Keck/DEIMOS spectroscopic study has targetted two fields in the
outskirts of the galaxy, located at $\sim 9$~kpc along the major axis
(McConnachie et al. 2006).  Although the dominant kinematic component
in these fields is rotating, there is tentative evidence for an
additional low-level metal-poor ([Fe/H]$\sim -1.5$) component centered
at the systemic velocity (see also Smecker-Hane, these proceedings).
While more work is needed to confirm this detection, the estimated
velocity dispersion ($\sigma \sim 50$~km/s) supports an association
with M33's true stellar halo.

\begin{figure}[ht!]
\centerline{\psfig{file=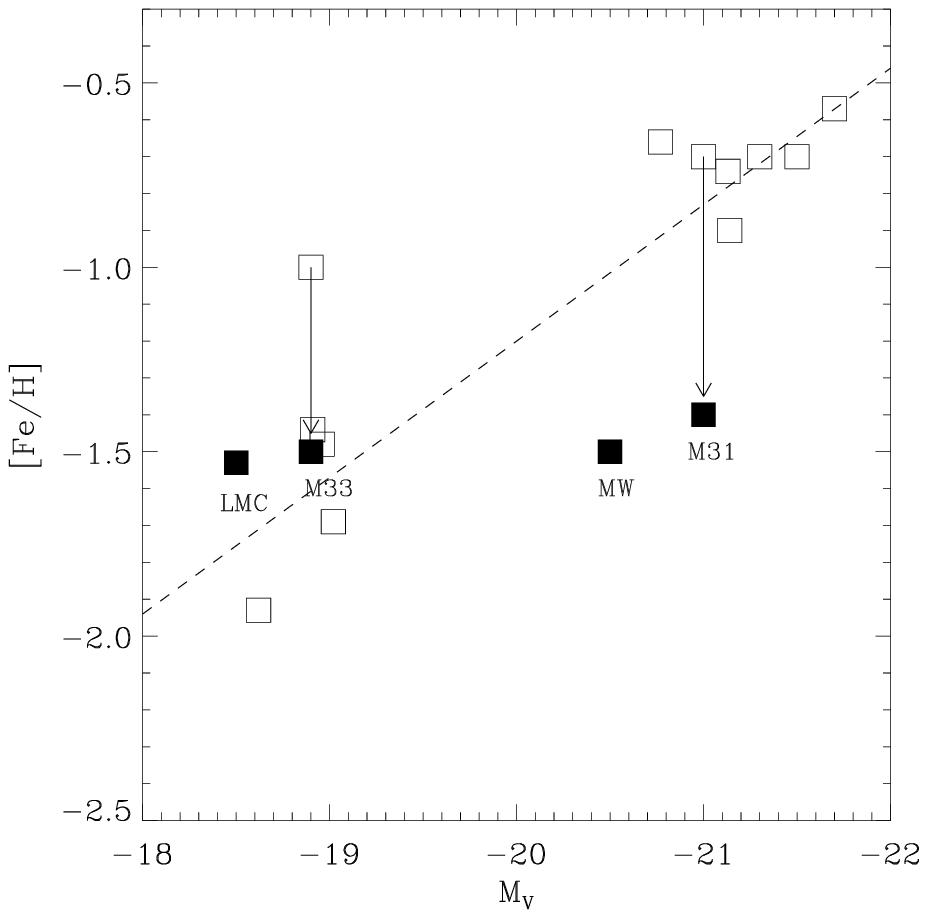,height=7cm,clip=}}
\caption{The mean metallicity of stellar halo populations plotted
against host galaxy absolute magnitude.  Open squares are taken from 
Mouhcine et al. (2005) and are derived from the colour of the RGB in
fields projected at some distance from the galaxy centre. The
solid squares represent spectroscopic metallicity estimates for
kinematically-selected halo stars, as described in the text.  Arrows
indicate substantial recent revisions to the metallicities of the
stellar halo populations in M31 and M33.}
\end{figure}

\subsection{Stellar Halos and Host Galaxies}

Based on the RGB colours at projected distances of 2-13~kpc along the
minor axes of 8 nearby spirals, Mouhcine et al. (2005) found evidence
for a correlation between stellar halo metallicity and parent galaxy
luminosity.  Specifically, they found that more luminous galaxies
possessed more metal-rich halos (and with a higher metallicity
spread). The existence (or otherwise) of such a correlation would
place extremely interesting constraints on the galaxy assembly
process.  Mouhcine et al. (2005) found the Milky Way to lie more than
1 dex off their observed relation; somewhat disconcertingly, they
suggested that the Milky Way halo may be more ``the exception than the
rule'' for large spiral galaxies.

Figure 4 presents the data from Mouhcine et al. (2005) along with our
new spectroscopic measurements of the stellar halo metallicity in M31
and M33. The metallicity of the Milky Way halo is also shown, as well
a recent spectroscopic metallicity derived for the high velocity
dispersion RR Lyrae component in the LMC (Borissova et al. 2006).
With the addition of these new data points, the correlation observed
between stellar halo metallicity and host galaxy luminosity is
substantially diminished. Indeed, if one focuses solely on
spectroscopic metallicities that have been determined for
kinematically-selected halo stars, there is no trend whatsoever
between metallicity and host luminosity.

\section{Future Outlook}
Quantitative study of the faint outskirts of galaxies provides
important insight into the galaxy assembly process. Much recent work
has focused on our nearest large neighbours, M31 and M33. While both
appear to show evidence for metal-poor, pressure-supported stellar
halos (similar to that of the Milky Way), only M31 shows evidence for
recent accretion.  Globular cluster populations offer additional clues
to the formation and evolution of M31.  This galaxy provides one of
the very few systems in which we can infer the assembly history from
both resolved field stars and globular clusters; it will be of great
importance to determine if a unified picture emerges from these rather
different approaches.

In order to put our Local Group results in context, it is necessary to
establish the properties of a larger sample of galaxies, spanning a
range in both host galaxy luminosity and Hubble type. With 8-m class
telescopes, it is possible to obtain wide-field maps of the RGB
populations in galaxies to distances of $\la 5$~Mpc;
with HST, this work can feasibly be extended to $\ga 10$~Mpc.
However, in this latter case, one must be mindful of the vagaries of
inferring global properties from small field-of-view studies of
components which have large angular extents on the sky. Spectroscopic
characterization of resolved RGB populations beyond the Local Group is
far harder and must await the arrival of 30-m class telescopes.

\acknowledgements 
I thank the organizers for a very enjoyable meeting.
Scott Chapman, Daniel Faria, Rodrigo Ibata, Mike Irwin, Avon Huxor,
Rachel Johnson, Kathryn Johnston, Geraint Lewis, Dougal Mackey,
Nicolas Martin, Alan McConnachie, Jenny Richardson and Nial Tanvir are
thanked for their collaboration. Support from a Marie Curie Excellence
Grant is acknowledged.


\end{document}